\documentclass[oneside, a4paper, onecolumn, 11pt]{article}
\usepackage[left=2.5cm,top=2.5cm,bottom=2.5cm,right=2.5cm]{geometry}
\usepackage{lineno}
\usepackage{amsmath}
\usepackage{graphicx}
\usepackage{fancyhdr}
\usepackage{calc}
\usepackage{amssymb}
\usepackage{lipsum}
\usepackage{titlesec}
\usepackage{color}
\usepackage{booktabs}
\usepackage[super,sort&compress,comma]{natbib}
\usepackage{hyperref}
\usepackage{url}
\hypersetup{
	colorlinks=true,
	linkcolor=black,
	filecolor=gray,      
	urlcolor=blue,
	citecolor=black,
}
\usepackage{setspace}
\usepackage{amsfonts}
\usepackage{commath}
\usepackage[innercaption]{sidecap}
\usepackage[framemethod=TikZ]{mdframed}
\usepackage{hyperref}
\usepackage{parskip}
\usepackage{pifont}
\usepackage[labelfont=bf]{caption}
\newcommand{\Rom}[1]{\expandafter\@slowromancap\romannumeral #1@}
\makeatother
{ \normalfont}

\expandafter\def\expandafter\normalsize\expandafter{%
	\normalsize
	\setlength\abovedisplayskip{0pt}
	\setlength\belowdisplayskip{5pt}
	\setlength\abovedisplayshortskip{0pt}
	\setlength\belowdisplayshortskip{5pt}
}

\definecolor{Gray}{gray}{0.75}
\definecolor{changecolour}{rgb}{0, 0, 0.8}

\newmdenv[backgroundcolor=Gray, leftmargin = 0pt, rightmargin = 0pt, linewidth = 0pt, roundcorner = 2 pt, innerleftmargin=5pt, innerrightmargin=5pt, innertopmargin=5pt, innerbottommargin=5pt]{Frame}

\let\oldequation\equation
\let\oldendequation\endequation
\renewenvironment{equation}{\linenomathNonumbers\oldequation}{\oldendequation\endlinenomath}

\let\oldalign\align
\let\oldendalign\endalign
\renewenvironment{align}{\linenomathNonumbers\oldalign}{\oldendalign\endlinenomath}

\let\oldgather\gather
\let\oldendgather\endgather
\renewenvironment{gather}{\linenomathNonumbers\oldgather}{\oldendgather\endlinenomath}

\begin{document}
	
	\newcommand{\kk}{\langle k \rangle}
	\newcommand{\kkk}{\langle k^2 \rangle}
	\newcommand{\er}{Erd\H{o}s-R\'{e}nyi}
	\newcommand{\red}{\color{red}\footnotesize}
	\newcommand{\blue}[1]{{\color{blue} #1}}
	\newcommand{\subfigimg}[3][,]{%
		
		\setbox1=\hbox{\includegraphics[#1]{#3}}
		\leavevmode\rlap{\usebox1}
		\rlap{\hspace*{30pt}\raisebox{\dimexpr\ht1-2\baselineskip}{#2}}
		\phantom{\usebox1}
	}
	
	\linespread{1.4}
	
	\begin{center}
		
		{\LARGE \textbf{Emergence of cooperation promoted by higher-order strategy updates}}
		
		\vspace{2mm}
		Dini Wang$^{1,2}$, Peng Yi$^{1,2*}$, Yiguang Hong$^{1,2}$, Jie Chen$^{1,2}$ and Gang Yan$^{2,3*}$
	\end{center}
	
	\small{
		\begin{enumerate}
			\item
			\textit{College of Electronic and Information Engineering, Tongji University, Shanghai, 201804, P. R. China}
			\item
			\textit{Shanghai Research Institute for Intelligent Autonomous Systems, National Key Laboratory of Autonomous Intelligent Unmanned Systems, MOE Frontiers Science Center for Intelligent Autonomous Systems, and Shanghai Key Laboratory of Intelligent Autonomous Systems, Tongji University, Shanghai, 201210, P. R. China}
			\item
			\textit{School of Physical Science and Engineering, Tongji University, Shanghai, 200092, P. R. China}
		\end{enumerate}
	}
	
	\vspace{5mm}
	\noindent
	\textbf{Cooperation is fundamental to human societies, and the interaction structure among individuals profoundly shapes its emergence and evolution. In real-world scenarios, cooperation prevails in multi-group (higher-order) populations, beyond just dyadic behaviors. Despite recent studies on group dilemmas in higher-order networks, the exploration of cooperation driven by higher-order strategy updates remains limited due to the intricacy and indivisibility of group-wise interactions. Here we investigate four categories of higher-order mechanisms for strategy updates in public goods games and establish their mathematical conditions for the emergence of cooperation. 
	Such conditions uncover the impact of both higher-order strategy updates and network properties on evolutionary outcomes, notably highlighting the enhancement of cooperation by overlaps between groups.
	Interestingly, we discover that the strategical mechanism alternating optimality and randomness -- selecting an outstanding group and then imitating a random individual within this group -- can prominently promote cooperation. Our analyses further unveil that, compared to pairwise interactions, higher-order strategy updates generally improve cooperation in most higher-order networks.
	These findings underscore the pivotal role of higher-order strategy updates in fostering collective cooperation in complex social systems.
	}

	\vspace{5mm}
	\textbf{\large Introduction}
	
	Cooperation emerges in nature, permeates evolutionary processes, and provides a strong foundation for human prosperity \cite{Hamilton1964-01, Hamilton1964-02, robert1981, nowak2011}. However, it raises the fundamental question: why would an individual sacrifice personal gain to support potential rivals in a competitive struggle? To understand the complex balance between cooperation and defection, evolutionary game theory offers valuable insights. It serves as a profound framework to describe the dynamic interactions in societal and economic behaviors \cite{hofbauer1998, nowak2006, press2012}. While natural selection generally favors defectors in well-mixed populations \cite{hofbauer1979, nowak2004-emergence}, the structure of a population -- defining the scope of individual interactions -- can significantly influence evolutionary outcomes \cite{nowak1992, debarre2014}. Networks, where nodes (representing individuals) are connected by edges (indicating interactions), are a key tool for exploring complex structured systems \cite{erez2005, li2020, yi2022survey, lei2020online, su2022, gao2022autonomous, chen2023outlearning, gao2024learning}; And network reciprocity has been proposed as a fundamental mechanism to address the cooperative dilemma, suggesting that certain population structures can foster local clusters of cooperators capable of resisting exploitation by defectors \cite{nowak1992, nowak2006, nowak2011}. Extensive research has investigated how network structures affect the evolution of cooperation through simulations \cite{nowak1992, santos2005, wes2014}, approximations \cite{ohtsuki2006}, and analytical solutions \cite{allen2017, alex2020}. These studies highlight the critical role of pairwise interactions in promoting cooperative behavior.
	
	Living systems frequently form multiple groups where members collaborate for the common good \cite{austin2016, lambiotte2019networks, battiston2021}. These groups can be defined by genetic relationships, social circles, local communities, or professional affiliations. Importantly, these groups often overlap, meaning individuals may belong to several groups simultaneously, thereby connecting groups through shared members \cite{ahn2010link}. Such overlapping memberships suggest the presence of an inherent higher-order network in real populations, where interactions occur not only within individual groups but also through the connections between groups \cite{Grilli2017, benson2018, petri2019, zhang2023, sarker2024}. The widespread prevalence of cooperation extends beyond dyadic interactions \cite{santos2008, tarnita2009pnas, gokhale2014}, raising the question: Can the dynamics of cooperative behavior in higher-order populations be fully explained by evolutionary games based on pairwise networks? Previous studies indicate that higher-order interactions lead to cooperative principles distinct from those observed in pairwise scenarios \cite{kumar2021evolution, guo2021, sheng2024}, suggesting that evolutionary games on higher-order networks warrant separate consideration. Despite recent advances in exploring various group social dilemmas within higher-order networks \cite{unai2021, andrea2021, sheng2024, civilini2024}, the potential of indivisible group-wise interactions to produce polyadic strategy updates remains not fully understood.
	
	The strategy updates in evolutionary games on higher-order networks fundamentally differ from those on pairwise networks. In each round of a game on a pairwise network, a player compares its own fitness with that of its neighbors. Based on this comparison, the player decides whether to adopt the strategy (i.e., cooperating or defecting) of the high-fitness neighbor or to retain its own strategy. In contrast, on a higher-order network, each player belongs to multiple groups, so the strategy update consists of two stages: selecting a group it belongs to and then selecting an individual in that group to imitate. Since both the group as a whole and the individuals within it retain their fitness, the player has numerous options in the two-stage selection process. Specifically, the first stage can be group-biased or not, while the second stage can be individual-biased or not.
	
	A player is group-biased if it prefer to select a group with highest fitness; not group-biased if it selects a group randomly. Similarly, a play is individual-biased if it prefer to imitate a high-fitness individual within the selected group. Combinations of these two-stage selection preferences could reflect four types of human personalities: aggressive (group-and-individual-biased), open-minded (group-biased), myopic (individual-biased), and passive (non-biased). An aggressive person tends to imitate a well-performing individual within an outstanding group, while an open-minded person interacts with and learns from an outstanding group as a whole. A myopic individual imitates a more skilled peer within a randomly chosen group, possibly due to an inability to evaluate all the groups they belong to. Passive individuals are willing to make random selections at both stages. These personalities are emblematic and pervasive across complex human societies. Accordingly, the four categories of selection preferences comprehensively capture the distinctive features of higher-order strategy updates. While a specific higher-order update has been explored in some group-wise structures \cite{unai2021}, general principles of cooperation driven by comparative higher-order updates have not yet been elucidated.
	
	Motivated by the above, our work provides a mathematical framework for the systematic analysis of higher-order updates that drive the strategy evolution of public goods games in higher-order networked populations. We first quantify the four categories of selection preferences as five specific higher-order update mechanisms, drawing analogies to pairwise rules. The aggressive type is further refined into two mechanisms by integrating death-birth (DB) and imitation (IM) processes. Building on these mechanisms, we establish higher-order random walk patterns to facilitate the application of coalescent theory. By integrating the individual-based mean-field approach with this framework, we derive analytical conditions for the  emergence of cooperation, applicable to any higher-order update mechanism and any higher-order network.
	
	Both the analytical solutions and experimental simulations reveal the strikingly different dynamics of the five update mechanisms. Notably, one mechanism, characterized by the imitation of a random individual in an outstanding group, demonstrates a particularly low threshold for the emergence of cooperation, indicating a strong ability to promote cooperative behavior. Our systematic analysis also explicitly reveals how higher-order topological factors independently influence the propensity for cooperation, encompassing network and correlation properties. Specifically, overlaps between groups, unique to higher-order networks, can promote cooperation. In comparison to pairwise updates, we find that higher-order strategy updates are generally more conducive to cooperation in a majority of higher-order networks. Thus, higher-order updates that drive the evolution of complex systems might provide a fundamental force for the thriving of cooperation in human societies.
	
	\vspace{5mm}
	\textbf{\large Results}
	
	\textbf{Higher-order networked game model and update mechanisms} \\
	We adopt the public goods game (PGG) as a fundamental model to study the group cooperative dilemma \cite{hardin1968, fowler2010, tavoni2011, hauser2019}.
	In each round of a PGG game in a  group of size $g$, the player serving as a cooperator (C) pays a cost $c$ for the public goods whereas the defector (D) makes no contribution. 
	Subsequently, all participants receive an equal share of the benefits, calculated by multiplying the total investment by the synergy factor $R$ with $1<R<g$ (see Figure~1a). 
	Without loss of generality, we fix the cost as $c=1$. 
	If there are $g_\text{C}$ cooperators in the group, the payoffs for both types of players in the PGG are given by:
	\begin{equation}
		u_\mathrm{C}=\frac{g_\mathrm{C}Rc}{g}-c,  \quad
		u_\mathrm{D}=\frac{g_\mathrm{C}Rc}{g}.
	\end{equation}
	Due to the dominant payoff, defection often appears more appealing from an individual perspective than cooperation. However, incorporating network structure into the population can potentially reverse the evolutionary dynamics of PGGs \cite{nowak1992, ohtsuki2006, unai2021}.
	
	We consider a population of $N$ interacting individuals, labeled as $\mathcal{N} = \{1, 2, \ldots, N \}$.
	To model the higher-order structure of this population, we use a hypergraph \cite{unai2021, andrea2021, martina2022}, as illustrated in Figure~1b. A hypergraph generalizes a pairwise graph by allowing hyperedges, which can connect more than two nodes, thus specifying higher-order interactions. That is, in a pairwise graph, an edge connects exactly two nodes, whereas in a hypergraph, a hyperedge can join multiple nodes. The order of a hyperedge $e$, denoted as $g_e$, is the number of nodes it connects and is a fundamental property of the hypergraph. For instance, in Figure~1b, the hyperedge $\beta$ includes four nodes, so $g_\beta = 4$. When the order of a hyperedge is two, as with the hyperedge $\sigma$ in Figure~1b, it degenerates to represent a pairwise interaction. A hypergraph is characterized by a set of such hyperedges, denoted as $\mathcal{E}$. Another key topological property of a hypergraph is the hyperdegree $k_i$, which represents the number of hyperedges adjacent to node $i$. For example, in Figure~1b, node 5 belongs to the hyperedges $\alpha$, $\beta$ and $\gamma$, so the hyperdegree of node 5 is $k_5=3$.
	
	In a hypergraph, strategy evolution can be described using a state vector $\mathbf{s} = (s_1, s_2, \ldots, s_N)^{\text{T}}$, where each individual $i$ is represented by $s_i = 1$ if they are a cooperator and $s_i = 0$ if they are a defector. The evolutionary process involves two key steps in each game round. First, players within a communal hyperedge engage in a game. Second, a randomly selected individual decides whether to change its strategy (from cooperator to defector, or vice versa) based on a higher-order update mechanism.
	
	For the first step, the local synergy factor $R_\alpha$ in the public goods game (PGG) of hyperedge $\alpha$ is normalized by the hyperedge's order to yield a global synergy factor $r = R_\alpha / g_\alpha$, where $g_\alpha$ is the order of the hyperedge. In other words, for a fixed $r$, $R_\alpha = r \cdot g_\alpha$ for each hyperedge $\alpha$. The payoff for any node $i$ is the average payoff over all of its adjacent hyperedges $e(i)$, quantified by:
	\begin{equation}
		u_i(\mathbf{s}) = \sum_{e(i) \in \mathcal{E}} \sum_{j \in e(i)} \frac{s_j}{k_i} r - s_i,
	\end{equation}
	where $r < 1$ induces a dilemma, while $r > 1$ resolves the dilemma. The final payoff of an individual depends on its hyperdegree and the number of cooperators within its adjacent hyperedges (Figure~1C). The fitness of an individual is then expressed as $f_i(\mathbf{s}) = \exp(\delta u_i(\mathbf{s}))$, where $\delta$ denotes the strength of selection \cite{nowak2004-emergence}. Here, $0 < \delta < 1$ measures how strongly the game outcome influences the individual's performance in reproduction or propagation \cite{ibsen2015}. For $\delta = 0$, the system undergoes neutral drift, and we focus on weak selection \cite{chen2013, allen2017, alex2020}, specifically the regime where $\delta \ll 1$. Based on this, the fitness of a group $e$ is defined as the average fitness of its members, calculated as $F_e(\mathbf{s}) = \sum_{i \in e} f_i(\mathbf{s}) / g_e$.
	
	For the second step, a player undergoing an update selects a specific neighbor (including themselves) for imitation or comparison through a two-stage selection process. Inspired by diverse personalities, we model four categories of two-stage selection preferences and quantify them into five distinct higher-order update mechanisms that drive the evolution of players' strategies. These higher-order update mechanisms are described as follows.
	
	The higher-order death-birth (HDB) mechanism, which inherits the aggressive trait from the dyadic death-birth (DB) update \cite{2004hauert}, involves the following steps: a random individual is chosen to die; the remaining individuals then select an adjacent hyperedge with a probability proportional to the group fitness, and within this hyperedge, a neighbor is selected with a probability proportional to the individual's fitness. The selected individual then replaces the vacant position, completing the strategy update under HDB. 
	
	The higher-order imitation mechanism (HIM) is similar to HDB, but with a key difference: in HIM, the individual to be replaced is included in both the group payoff calculation and the selection process, whereas in HDB, it is not. Both HDB and HIM represent group-and-individual-biased selection (Figure~1d). 
	
	In contrast, the group-mutual-comparison (GMC) and group-inner-comparison (GIC) mechanisms are characterized by group-biased and individual-biased selections, representing open-minded and myopic, respectively. GMC selects a group based on fitness evaluation, while GIC selects an individual within a specific group based on fitness evaluation (Figure~1e-f). 
	
	Finally, the higher-order pair-comparison (HPC) mechanism represents non-biased selection, where both the adjacent hyperedge and the peer within the hyperedge are chosen uniformly at random. The player undergoing update then compares its fitness with that of the randomly chosen peer to decide whether to adopt the peer's strategy (Figure~1g). 
	
	Essentially, these five update mechanisms differ in their degree of greediness during the decision-making process for selecting an adjacent hyperedge and an individual within that hyperedge. For detailed descriptions, see Extended Table 1.

	\textbf{Emergence of cooperation under five higher-order update mechanisms} \\
	Our primary objective is to assess whether cooperation is advantageous in a specific hypergraph driven by higher-order updates, and if so, to what extent. In the absence of mutation, the evolutionary system ultimately reaches an absorbing state where all individuals are either cooperators or defectors after a sufficient number of game rounds \cite{allen2014mathe}. The fixation probability of cooperation, denoted as $\rho_\mathrm{C}$, represents the likelihood that a randomly introduced cooperator can convert the entire population from defection to cooperation, while $\rho_\mathrm{D}$ is defined similarly for defectors. In the case of neutral drift ($\delta = 0$), both fixation probabilities are $1/N$, where $N$ is the number of nodes in the system. Therefore, if $\rho_\mathrm{C} > 1/N > \rho_\mathrm{D}$, the evolutionary mechanism favors cooperation over defection \cite{drew2006}. 
	
	Consequently, our investigation focuses on determining the critical synergy factor for cooperation -- specifically, the smallest value of synergy factor $r^*$ for which $\rho_\mathrm{C}$ exceeds the baseline of $1/N$. If $r^*$ is less than one for a structured population under a given update mechanism, such hypergraph can be considered to foster cooperation, as the threshold for cooperation is lower compared to an isolated public goods game (PGG), and vice versa. A smaller critical synergy factor will support a greater proliferation of cooperators when $1/g < r^* < 1$. Moreover, we aim to analytically explore how the critical synergy factor $r^*$ is related to hypergraph properties under different update mechanisms, thereby validating the principle of network reciprocity in hypergraphs.
	
	Mathematically, the evolutionary process is essentially a complex, non-stationary Markovian chain driven by game fitness \cite{Gomez2010}. To simplify the interactions among individuals in the population, we use probabilistic state transformations and adopt an individual-based mean-field approach, which assumes negligible dynamical correlations between the strategies of neighbors \cite{goltsev2012, Wang2017}. Under this assumption, weak-selection perturbations of fixation probabilities reduce the problem of identifying cooperation triggers to a probabilistic state-dependent formula \cite{asava2001, tan2015} in the case of neutral drift ($\delta=0$). Quantitative analysis of strategy assortments in this neutral scenario employs coalescence theory, which traces the lineage of an initial mutant by looking backward at ancestral states \cite{kingman1982, chen2013}.
	
	Guided by higher-order update mechanisms, we develop two patterns of higher-order random walks to effectively integrate coalescence theory into the hypergraph context. This theoretical framework provides explicit critical synergy factors for public goods games (PGG) on any hypergraph and under any higher-order update mechanism, as detailed in Table 1 (see Supplementary Information Sections 3 - 6 for the mathematical derivations). These results establish the necessary and sufficient conditions for the emergence of cooperation under higher-order update mechanisms.

	\renewcommand{\arraystretch}{2}
	\begin{table}[h!]
		\centering
		\begin{tabular}{cc}
			\hline
			Update mechanism & Critical synergy factor, $r^*$  \\
			\midrule
			HDB & $\dfrac{{N-2\eta_k} }{{N + N/ \langle{k}\rangle + N\widetilde{\theta} -2\zeta} }$  \\[5pt]
			HIM & $\dfrac{{N+N/  \langle{g}\rangle  -2 \eta_k} }{ {N+N/   \langle{k} \rangle+N\widehat{\theta}-2\zeta} }$  \\[5pt]
			GMC & $\dfrac{{N / \langle{g}\rangle  -\eta_k} }{ N/\langle{k}\rangle + { N\widehat{\theta}-\zeta}}$ \\[5pt]
			GIC & $\dfrac{{N-\eta_k} }{{N-\zeta}}$ \\[5pt]
			HPC & $\dfrac{{N-\eta_k} }{{N-\zeta}}$ \\[5pt]
			\bottomrule
		\end{tabular}
		\caption{Critical synergy factor for five higher-order update mechanisms. Here $N$ is the population size; $\langle k \rangle$ is the average hyperdegree of the node; $\langle g \rangle$ is the average order of the hyperedge; $\eta_k$ denotes the hyperdegree heterogeneity; $\zeta$ indicates the assortativity coefficient between the node's hyperdegree and the hyperedge's order; $\widetilde{\theta}$ and $\widehat{\theta}$ indicate the overlap strength between hyperedges in analogous manners};
	\end{table}
	
	For such complex evolutionary processes of public goods games (PGGs) on hypergraphs, our findings in Table 1 surprisingly unveil that the critical thresholds $r^\ast$ required for cooperation are generally determined by two types of hypergraph topological factors: basic properties and correlation properties. The basic properties include the average order, $\langle g \rangle = \sum_{\alpha \in \mathcal{E}} g_\alpha / E$, which measures the expected size of group-wise interactions; the average hyperdegree, $\langle k \rangle = \sum_{i \in \mathcal{N}} k_i / N$, which quantifies the connection density of the global population; and hyperdegree heterogeneity, defined as $\eta_k = \langle k^2 \rangle / \langle k \rangle^2$, where $\langle k^2 \rangle = \sum_{i \in \mathcal{N}} k_i^2 / N$ represents the second moment of the hyperdegree distribution. These factors extend the concepts from pairwise networks. In addition to these basic properties, hypergraphs exhibit unique correlation properties, including the overlap strength between hyperedges, denoted $\widehat{\theta}$ or $\widetilde{\theta}$, which quantifies the extent of node overlap between adjacent hyperedges; the assortativity coefficient, $\zeta$, which measures the tendency of nodes with similar hyperdegrees to connect with hyperedges of similar order. Detailed definitions of these correlation properties are provided in the section Methods. The subsequent two subsections will explore how both basic and correlation properties influence the emergence of cooperation in hypergraphs.
	
	To validate our analytical results, we perform Monte Carlo simulations on random hypergraphs with both homogeneous and heterogeneous orders and hyperdegrees, encompassing a wide range of hypergraph structures. As shown in Figure~2, the experimental results, represented by scatter plots, align closely with the analytical solutions indicated by arrows. Furthermore, we test the analytical predictions on empirical higher-order networks, demonstrating accurate predictions of the real synergy factors required for cooperation (see Table S2 in the Supplementary Information). 
	
	Our results reveal fundamentally different impacts of various higher-order update mechanisms on the promotion of cooperation. As shown in Figure~2, the critical synergy factors for the HPC and GIC mechanisms consistently exceed one. This outcome is confirmed by theoretical analysis: since $\zeta > \eta_k$ for any hypergraph, the numerator in the critical synergy factors for HPC and GIC is always larger than the denominator. Therefore, both HPC and GIC mechanisms hinder cooperation across all hypergraphs, consistent with the traditional pairwise comparison update in network-based games.
	
	In contrast, the critical synergy factor for the GMC mechanism is significantly smaller than one and lower than those for the HDB and HIM mechanisms. This comparison is particularly noteworthy because HDB and HIM mechanisms reflect aggressive strategy updates -- selecting an outstanding group and imitating a high-fitness individual within that group -- while the GMC mechanism represents a combination of optimality and randomness -- selecting an outstanding group and then imitating a random individual within that group. Thus, our findings suggest that indiscriminate interactions among members of an outstanding group, rather than excessive greediness, are crucial for fostering the spread of cooperation within societal systems.
	
	\textbf{Impact of hyperdegree and order on cooperation} \\
	Here we quantitatively investigate how hyperdegree, order, and their heterogeneities influence the emergence and evolution of cooperation in large populations. This analysis is conducted by taking partial derivatives of the critical synergy factors with respect to these parameters (details are provided in Supplementary Information Section 7).
	
	Since HPC and GIC mechanisms clearly inhibit cooperation, we focus on the effects of GMC, HDB, and HIM mechanisms. We first examine homogeneous hypergraphs, where each node has the same hyperdegree and each hyperedge has the same order (Figure~2a). In such homogeneous hypergraphs, a smaller hyperdegree is associated with a lower critical synergy factor required for triggering cooperation under GMC, HDB, and HIM mechanisms (Figure~3a). This trend mirrors the scenario in pairwise updates, where a lower degree corresponds to a reduced threshold for cooperation \cite{ohtsuki2006}. This relationship also aligns with Hamilton's rule, which suggests that fewer social connections lead to stronger ties within those connections \cite{Hamilton1964-01, Hamilton1964-02}.
	
	The effect of homogeneous order on cooperation varies across different higher-order update mechanisms (Figure~3b). For the GMC mechanism, the critical synergy factor decreases as the hyperedge's order increases, indicating that larger hyperedges tend to promote cooperation. Notably, when the order approaches the population size, the GMC mechanism effectively becomes a dyadic update under neutral drift, where the synergy factor does not influence the evolutionary outcome. Additionally, for $g=2$ (pairwise interactions), the critical synergy factors of GMC and HDB are identical, a result confirmed by our theoretical analysis. In contrast to GMC, a smaller order generally facilitates cooperation in HDB and HIM mechanisms because the order reflects the number of neighboring alternatives in the second stage of selection. However, the threshold for the HIM mechanism shows a slight decrease with very small orders. This is because smaller hyperedges increase the likelihood of self-selection during imitation, which may inhibit the spread of cooperation.
	
	Despite these insights, assuming uniformity among individuals is an idealization that does not fully capture real-world heterogeneity. Thus we further explore the effect of the heterogeneity in hyperdegree and order. The results reveal that weaker heterogeneity tends to foster cooperative behaviors (Figure~3c-d). Notably, hyperdegree heterogeneity has a minimal impact on evolutionary outcomes under the GMC mechanism. Consequently, variations in order introduce rich and intriguing dynamics in higher-order interactions, contrasting with the fixed order of two in pairwise networks.

	\textbf{Overlap between hyperedges promotes cooperation}\\
	The overlap between hyperedges is a fundamental characteristic of hypergraphs. To investigate how this overlap impacts the emergence of cooperation, we use homogeneous hypergraphs to control for complex topological variations. Based on the definition of overlap strength $\widehat{\theta}$ (see section Methods), we quantify the rescaled overlap strength for homogeneous hypergraphs as follows:
	\begin{equation}                    
		C_{\mathrm{ovl}} = \frac{\sum_{\alpha, \beta \in \mathcal{E}, \alpha \neq \beta} |\alpha \cap \beta|^2}{Nk(k-1)g}.
		\label{overlap-homo}
	\end{equation}
	Here, $|\alpha \cap \beta|$ denotes the number of shared nodes between hyperedges $\alpha$ and $\beta$, and the denominator normalizes this measure. When multiple hyperedges share a large number of nodes, the overlap strength increases quadratically. Mathematically, the overlap strength $C_{\mathrm{ovl}}$ ranges from $1/g$ to $1$ -- where $1/g$ is the minimum overlap strength when each pair of adjacent hyperedges overlaps over only one node, and $1$ is the maximum value when the overlap equals the hyperedges' order.
	
	Take the GMC mechanism for an example, its critical synergy factor $r^\ast$ related to overlap strength can be recast as:
	\begin{equation}
		r^* = \frac{{\frac{N}{g} - 1}}{\frac{N}{k} + N\left(1 - \frac{1}{k}\right)C_{\text{ovl}} - g}.
	\end{equation}
	This formula indicates that a higher $C_{\text{ovl}}$ corresponds to a lower critical synergy factor required for the emergence of cooperation. We numerically validate this theoretical prediction using two groups of homogeneous hypergraphs of large scale, shown in Figure~4a-c and Figure~4e-g, each with identical hyperdegree and order but differing overlap patterns.
	
	As exhibited in Figure~4d and Figure~4h, the critical synergy factor decreases with increasing overlap strength. This result suggests that larger overlaps among hyperedges lead to more frequent interactions, as individuals can engage with others through multiple groups. Such enhanced connectivity facilitates the spread of cooperation throughout the population under the GMC update mechanism.

	\textbf{Intuition for distinct higher-order strategy updates}\\
	To intuitively understand how different higher-order strategy updates influence the evolution of cooperation, consider the challenge of a single cooperator emerging in a population of defectors. This situation can be reframed as the problem of how defectors might convert to cooperation under various higher-order update mechanisms.
	
	From the perspective of a defector, the two-stage process of selecting a target individual for imitation or comparison can be described as follows: Firstly, group selection can either be fitness-biased (or payoff-biased) or completely neutral, as illustrated in Figure~5a. Groups with more cooperators generally yield higher group payoffs, making group-biased selection inclined toward these cooperator-rich groups. In contrast, group-neutral selection treats all groups equally. Thus, group-biased selection methods, such as GMC, HDB, and HIM, create favorable conditions for cooperators to reproduce, establishing a solid foundation for higher-order updates. Secondly, individual selection, as shown in Figure~5b, can also be either biased or neutral. In this context, defectors often outperform cooperators within a given group, leading to greater reproductive success for defectors. Therefore, individual-neutral selection, as seen in GMC and HPC, tends to promote cooperative behaviors. The key difference lies in the final strategy update procedure. While direct imitation of a cooperator can promote cooperation, comparison with a higher-payoff individual (as in HPC) may reinforce the existing defecting strategy, reducing the likelihood of cooperation.
	
	In summary, group-biased selection is crucial for setting the stage for evolutionary success, while individual-neutral selection helps further promote cooperation. However, comparison processes, such as those used in HPC, tend to reinforce defective behaviors rather than cooperation. Thus, GMC, which combines group-biased selection with individual-neutral selection, emerges as the most effective update mechanism. In contrast, mechanisms like GIC and HPC struggle to drive the population towards cooperation due to their lack of emphasis on selecting well-performing groups. These findings align well with both analytical results and experimental simulations, providing valuable insights into the dynamics of higher-order updates.

	\textbf{Comparison between higher-order and pairwise updates}\\
	To deepen our understanding of higher-order strategy updates, we compare them with classic pairwise updates within the same game setting. The key distinction lies in the selection process for imitation or comparison. In higher-order updates, the to-be-replaced player undergoes a two-stage selection process: first, group selection, followed by individual selection (Figure~6a). In contrast, pairwise updates involve a straightforward selection of an individual (including oneself) from all adjacent nodes (Figure~6b).
	
	When the hyperedge order $g = 2$, some higher-order updates reduce to pairwise updates. For instance, HDB reduces to DB and HPC reduces to PC. Theoretical values for critical synergy factors under DB and PC mechanisms are derived and compared (see Supplementary Information Section 8). We show the differences in synergy factors required for cooperation between higher-order and pairwise updates using empirical data from higher-order populations and small-scale hypergraphs.
	
	To analyze multi-group behaviors in real-world contexts, we examine the congress bills network in the U.S. This network includes 57 nodes representing congresspersons and 108 hyperedges representing bill sponsors and co-sponsors (Figure~6c). We numerically simulate the evolutionary process of public good game for both higher-order and pairwise updates until the system reaches a fixed state. Our simulations confirm that theoretical predictions accurately reflect experimental outcomes. We find that both HDB and HPC mechanisms have significantly lower critical synergy factors compared to their pairwise counterparts (Figure~6d), indicating that higher-order interactions generally support the spread of cooperation.
	
	Additionally, we analyze various small-scale higher-order structures of size seven, derived from cliques in simple graphs. We first study the propensity for cooperation in straightforward pairwise interactions and then compare these results with higher-order updates. As shown in Figure~6e, the critical synergy factors for higher-order updates generally encompass those for pairwise updates. Notably, significant differences arise with large hyperedges and strong overlaps, reflecting the unique aspects of higher-order networks. This analysis provides new insights into the organization of higher-order structures, reinforcing that higher-order updates enhance the prospects for cooperation.

	\vspace{5mm}
	\textbf{\large Discussion}\\
	A substantial body of research has demonstrated that the population structure of autonomous individuals can significantly influence, and even alter, the trajectory of evolving cooperation. This structure extends beyond simple peer-to-peer interactions to encompass the more complex multi-group (higher-order) dynamics prevalent in human societies. In higher-order populations, both group-level and individual-level factors collectively influence an individual’s strategy update throughout evolution. Essentially, higher-order updates involve a two-stage selection process: first selecting a neighboring group and then selecting an individual within that group. Inspired by the pluralistic nature of social interactions, we categorize higher-order updates into four types and qualitatively describe their two-stage selection preferences. Understanding how these higher-order updates impact interaction dynamics is crucial for explaining evolutionary processes in multi-group settings.
	
	We model higher-order populations using hypergraphs, where nodes (representing individuals) engage in public goods games (PGGs) on hyperedges and adjust their strategies based on various update mechanisms. We quantify higher-order strategy updates through five specific mechanisms with rigorous mathematical treatment. By deriving analytical cooperation thresholds for any hypergraph driven by these mechanisms, we reveal the impact of hypergraph properties on critical synergy factors required for the emergence of cooperation.
	
	Notably, the GMC mechanism significantly promotes the spread of cooperative behaviors, and overlaps between hyperedges generally facilitate cooperation. Numerical calculations on random and empirical hypergraphs demonstrate that higher-order updates are often more conducive to cooperation compared to dyadic updates. This paradigm deepens our understanding of evolutionary dynamics in complex social as well as biological systems.
	
	Our work provides novel insights into the evolution of cooperation, particularly highlighting the impact of the GMC mechanism. GMC, an open-minded update strategy, catalyzes cooperation by ensuring a lower critical synergy factor for PGGs. It emphasizes the selection of well-performing groups and treats individuals within the group uniformly. Specifically, GMC's critical synergy factors required for cooperation can be up to a hundredfold lower than those of other mechanisms. Despite the minimal direct benefits to individual players in PGGs, GMC effectively converts defectors into cooperators across the population. This suggests that alternating between optimality and randomness, rather than excessive greediness, can promote the propagation of cooperative behaviors across populations with groups representing diverse interests or cultures.
	
	Another significant insight is that overlaps between different groups, unique to higher-order networks, can substantially promote cooperation under certain higher-order update mechanisms. Even with fixed order and hyperdegree, varying overlap strengths in higher-order networks reflect the richness of multi-group interactions. Our theoretical analysis confirms that greater overlap strength -- resulting in more frequent interactions and stronger social bonds --can effectively lower the critical synergy factor required for cooperation. Furthermore, overlapping isolated populations can significantly reduce the barriers to cooperation (Supplementary Information 9), highlighting the potential for cooperation to thrive even in otherwise isolated settings. This insight has implications for cross-national interactions, suggesting that increased multicultural acceptance and integration could foster international cooperation.
	
	Despite the intriguing phenomena and implications, our study has several limitations. We assume a fixed higher-order structure for the population, even though traits typically evolve faster than social connections. Exploring variations in group memberships and their dynamic effects on prosocial behaviors could reveal new scientific discoveries. Additionally, our study focuses on a single update mechanism. Social domains may exhibit various cultural preferences or communication styles, warranting further research into the coexistence of multiple mechanisms and their effects. For instance, biased selection models, as opposed to weak probabilistic inclinations, present challenges for mathematical treatment. Future research should explore these aspects to advance our understanding of evolving cooperation.

	\vspace{5mm}
	\textbf{\large Methods}\\
	The model and mathematical methods are summarized in the following and the complete derivations are provided in Supplementary Information.
	
	\textbf{Hypergraph and higher-order random walk}\\
	The structure of a higher-order population can be represented as a connected hypergraph of size $N$, composed of a node set $\mathcal{N}$ and a hyperedge set $\mathcal{E}$. 
	To describe the affiliation of nodes and hyperedges, consider the incidence matrix $(b(i,e))_{i \in \mathcal{N}, e \in \mathcal{E}}$ where $b(i,e)=1$ indicates that node $i$ belongs to hyperedge $e$, and $b(i,e)=0$ inidcates no such belonging. The order denotes the size of a hyperedge, defined as $g_e = \sum_{i \in \mathcal{N}} b(i,e)$ for $e \in \mathcal{E}$; and the hyperedgree denotes the number of hyperedges  adjacent to a node, defined as $k_i = \sum_{e \in \mathcal{E}} b(i,e)$ for $i \in \mathcal{N}$.
	
	We explore two patterns of higher-order random walk to trace the original cooperative strategy on the hypergraph under five update mechanisms. We begin with the higher-order random walk with no self-loops, where a walker moves from a node $i$ to another node $j$ by first selecting an adjacent hyperedge and then selecting a neighbor within that hyperedge, with probability $\widetilde{p}_{ij}$. Similarly, the higher-order random walk with self-loops allows an individual to stay at their current position. The probability of moving from node $i$ to node $j$ in this case is defined as $\widehat{p}_{ij}$.
	
	We introduce a novel $(n,m)$-hop random walk pattern, representing a walk of $n$ hops without self-loops followed by $m$ hops with self-loops. The probability of an $(n,m)$-hop random walk from node $i$ to node $j$ is given by
	\begin{equation}
		p_{ij}^{(n,m)} = \sum_{x \in \mathcal{N}} \widetilde{p}_{ix}^{(n)} \widehat{p}_{xj}^{(m)}.
	\end{equation}
	Refer to Supplementary Information Section 1 for the details.
	
	\textbf{Modeling evolutionary game on hypergraph}\\
	For the public goods game on a hypergraph, each node can be either a cooperator or a defector. Let $\mathbf{s} = (s_1, s_2, \ldots, s_N)^{\text{T}}$ represent the state vector of the hypergraph, where $s_i = 1$ denotes a cooperator and $s_i = 0$ denotes a defector. Based on the state vector $\mathbf{s}$, each node $i$ has an individual fitness $f_i(\mathbf{s})$, and each hyperedge $e$ has a group fitness $F_e(\mathbf{s})$ in a given round of the game.
	
	To drive the evolution of the game, we describe five higher-order strategy updates based on two-stage selection preferences. Taking the GMC mechanism as an example, the update process for individual $i$ in the state $\mathbf{s}$ is divided into two stages: first, selecting a group that it belongs to $e$ with the probability
	\begin{equation}
		\mathrm{Pr} \left[i \rightarrow e\right]\left(\mathbf{s}\right)
		= \frac{b(i,e) {F}_{e}(\mathbf{s})}{\sum_{\alpha \in \mathcal{E}} b(i,\alpha) {F}_{\alpha }(\mathbf{s})},
	\end{equation}
	and second, imitating a neighbor $j$ $(j \neq i)$ within group $e$ with probability
	\begin{equation}
		\mathrm{Pr} \left[i \rightarrow j | i \rightarrow e \right]\left(\mathbf{s}\right)
		= \frac{b(j,e)}{g_e-1}.
	\end{equation}
	
	Combining both stages, the transition rate that $i$ adopts $j$'s strategy under the GMC mechanism is given by
	\begin{equation}
		\mathrm{Pr} \left[i \rightarrow j\right]\left(\mathbf{s}\right) 
		= \sum_{e \in \mathcal{E}} \mathrm{Pr} \left[i \rightarrow e\right]\left(\mathbf{s}\right) \cdot \mathrm{Pr} \left[i \rightarrow j | i \rightarrow e \right]\left(\mathbf{s}\right).
	\end{equation}
	
	Refer to Supplementary Information Sections 2 and 6 for the modeling under other higher-order update mechanisms.
	
	\textbf{State transformation in a probabilistic sense}\\
	To analyze strategy evolution, we introduce a probability vector for the system, $\mathbf{x} = (x_1, x_2, \ldots, x_N)^{\text{T}}$, which corresponds to the state vector $\mathbf{s}$. Here, $x_i = \mathrm{Pr}(s_i = 1)$ indicates the probability of node $i$ being a cooperator. Hence, $0 \leq x_i \leq 1$, where $x_i = 1$ for a cooperator and $x_i = 0$ for a defector.
	
	To describe the dynamic process, the state vector over time is given by $\mathbf{s}(t) = (s_1(t), s_2(t), \ldots, s_N(t))^{\text{T}}$, where $s_i(t)$ indicates the state of node $i$ at time $t$. Similarly, the probability vector over time is $\mathbf{x}(t) = (x_1(t), x_2(t), \ldots, x_N(t))^{\text{T}}$, where $x_i(t$ indicates the probability that node $i$ is a cooperator at time $t$.
	
	For an update process, the state of the hypergraph evolves according to the following formula:
	\begin{equation}
		\begin{aligned}
			&\mathbf{x}(t+1) = M(\mathbf{s}(t)) \cdot \mathbf{s}(t), \\
			&\mathbf{s}(t+1) = \mathcal{R}(\mathbf{x}(t+1)),
		\end{aligned}
		\label{prob model}
	\end{equation}
	where $M(\mathbf{s}(t))$ is a transition matrix based on the current state $\mathbf{s}(t)$. Each element $\left[M(\mathbf{s}(t))\right]_{ij}$ represents the strategy transition rate from $j$ to $i$, given by $\mathrm{Pr} \left[i \rightarrow j\right] (\mathbf{s}(t))$. The operator $\mathcal{R}(\cdot)$ is a realization of $\mathbf{x}(t+1)$. Specifically, $s_i(t+1)$ is determined to be 1 with the probability $x_i(t+1)$ or 0 with the probability $1-x_i(t+1)$ for $\in \mathcal{N}$. Thus, \(\mathbf{x}(t) = \mathbb{E}(\mathbf{s}(t))\) at each time step, and hereafter we consider the evolving states from a probabilistic perspective.
	
	\textbf{Derivation of critical synergy factor}\\
	Here, we briefly outline the derivation for the GMC update mechanism, with complete derivations for the other four mechanisms detailed in Supplementary Information Section 6. By neglecting dynamic correlations between the states of neighboring nodes, the individual-based mean-field approach yields the condition for cooperation under the GMC mechanism as:
	\begin{equation}
		\sum_{i \in \mathcal{N}} \pi_i \left\langle x_i \left( u_i^{(0,1)}(\mathbf{x}) - u_i^{(1,1)}(\mathbf{x}) \right) \right\rangle > 0,
	\end{equation}
	where \(\langle \cdot \rangle\) denotes averaging over state assortments across time scales under neutral drift and a fixed position for the initial mutant. Here, \(u_i^{(n,m)}(\mathbf{x})\) represents the expected payoff of \( (n,m) \)-hop neighbors of node \(i\) in the probabilistic state \(\mathbf{x}\), and \(\pi_i\) denotes the normalized hyperdegree of node \(i\).
	
	With neutral drift (\(\delta = 0\)), the transition matrix \(M(\mathbf{s}(t))\) in the time interval from [\ref{prob model}] can be described as a fixed stochastic matrix \(\Xi\), where each element \(\xi_{ij}\) depends solely on the network structure in which the trait evolves. Thus, the expected state of node \(i\) at time \(t\) can be traced back to the initial configuration, given by:
	\[
	x_i(t) = \sum_{u \in \mathcal{N}} \xi_{iu}^{(t)} x_u(0).
	\]
	Considering integral transformation and dislocation elimination, the critical synergy factor required to trigger cooperation is:
	\begin{equation}
		r^* = \frac{ \sum_{v \in \mathcal{N}} \pi_v \left( p_{vv}^{(0,1)} - \pi_v \right) }{ \sum_{i,v \in \mathcal{N}} \pi_v \left( p_{vi}^{(0,1)} - \pi_i \right) \sum_{e \in \mathcal{E}} \frac{b(i,e) b(v,e)}{k_i} }.
		\label{r-star}
	\end{equation}
	The equivalent and explicit form for the GMC mechanism is provided in Table 1. When the synergy factor \(r\) exceeds the threshold \(r^*\), selection favors cooperation over defection, leading to its expansion and fixation across the population.
	
	\textbf{Correlation properties of hypergraph}\\
	To investigate the impact of higher-order structural properties on the threshold for cooperation, we examine two key correlation properties of the hypergraph.
	
	The first property is the overlap strength between hyperedges, quantified by the following expressions:
	\begin{equation} \label{theta_hat}
		\widehat{\theta} = \frac{1}{N \langle k \rangle} \sum_{\alpha, \beta \in \mathcal{E}, \alpha \neq \beta} \sum_{i \in \alpha \cap \beta} \frac{|\alpha \cap \beta|}{k_i g_{\alpha}},
	\end{equation}
	and 
	\begin{equation} \label{theta_tilde}
		\widetilde{\theta} = \frac{1}{N \langle k \rangle} \sum_{\alpha, \beta \in \mathcal{E}, \alpha \neq \beta} \sum_{i \in \alpha \cap \beta} \frac{|\alpha \cap \beta| - 1}{k_i (g_{\alpha} - 1)},
	\end{equation}
	where $|\cdot|$ denotes the size of the set. We compare the overlap strength for hypergraphs with identical hyperdegree distributions and identical order distributions, ensuring consistent basic structural factors. Specifically, for homogeneous hypergraphs -- where all nodes have the same hyperdegree and all hyperedges have the same order -- the overlap strength in [\ref{overlap-homo}] is a rescaled version of $\widetilde{\theta}$.
	
	The second property is the assortativity coefficient between a node's hyperdegree and the hyperedges' order, given by
	\begin{equation}
		\zeta = \frac{1}{N \langle k \rangle^2} \sum_{\alpha \in \mathcal{E}} \sum_{i \in \alpha} g_{\alpha} k_i,
	\end{equation}
	where $\langle k \rangle$ is the average hyperdegree. Both of these correlation properties can significantly influence interaction dynamics thus play a crucial role in the emergence and evolution of cooperation.
	
	\vspace{5mm}
	\textbf{\large Data availability}: Several empirical hypergraphs are used in this study and the data can be accessed via https://github.com/arbenson/ScHoLP-Data.
	
	\vspace{5mm}
	\textbf{\large Code availability}: Code that supports the findings of this study is available in \textit{GitHub} via the link https://github.com/diniwang/EvolutionaryGame-Hypergraph.
	
	\vspace{5mm}
	\textbf{\large Acknowledgements}: This work was supported by the National Natural Science Foundation of China (Grants No. T2225022, No. 12161141016, No. 62088101, No. 72271187, and No. 62373283), Shanghai Municipal Science and Technology Major Project (Grant No. 2021SHZDZX0100), National Key Research and Development Program of China (Grant No. 2022YFA1004701), and the Fundamental Research Funds for the Central Universities.

	\vspace{5mm}
	\textbf{\large Author contributions}:
	GY and PY conceived the project; DW performed mathematical analysis and numerical simulations with help from PY and GY; GY, PY, YH and JC analyzed the results; DW, GY and PY wrote the paper with input from YH and JC. All authors contributed to all aspects of the project.
	
	\vspace{5mm}
	\textbf{\large Corresponding author}: Correspondence to Peng Yi (yipeng@tongji.edu.cn) or Gang Yan\\ (gyan@tongji.edu.cn).
	
	\vspace{5mm}
	\textbf{\large Competing interests}: The authors declare no competing interests.
	
	\clearpage
	
	\begin{figure*}[t!]
		\centering
		\includegraphics[width=\textwidth]{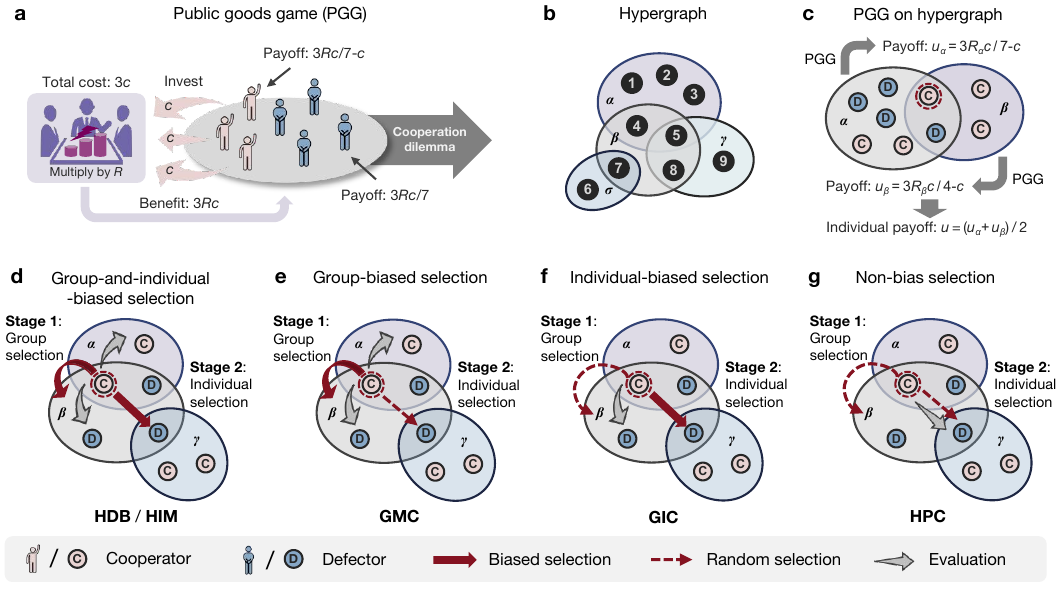}
		\caption{Public goods game (PGG) on a hypergraph and four categories of higher-order update mechanisms. 
			\textbf{a}, In a PGG, each player is either a cooperator (pink) or a defector (blue). Every cooperator invest a cost \(c\), while defectors do not. The total investment is multiplied by the synergy factor \(R\) to produce a benefit $3cR$ shared equally among all players, resulting in a cooperative dilemma. 
			\textbf{b}, A nine-node hypergraph with four hyperedges ($\alpha$: $\{1,2,3,4,5\}$, $\beta$: $\{4,5,7,8\}$, $\gamma$: $\{6,7\}$, and $\sigma$: $\{5,8,9\}$).
			\textbf{c}, Players on the hypergraph participate in PGGs within the hyperedges. For example, the focal node (circled in red) belongs to two hyperedges, $\alpha$ and $\beta$, hence it participates two games and obtains the payoffs averaged across these two games. 
			\textbf{d}-\textbf{g}, Four categories of two-stage updates: \textbf{d}, Group-and-individual-biased (HDB for higher-order death-birth, HIM for higher-order imitation); \textbf{e}, Group-biased (GMC for group-mutual comparison); \textbf{f}, Individual-biased (GIC for group-inner comparison); and \textbf{g}, Non-biased (HPC for higher-order pair-comparison).
		}\label{fig:side}
	\end{figure*}
	
	\clearpage
	\begin{figure*}[t!]
		\centering
		\includegraphics[width=\textwidth]{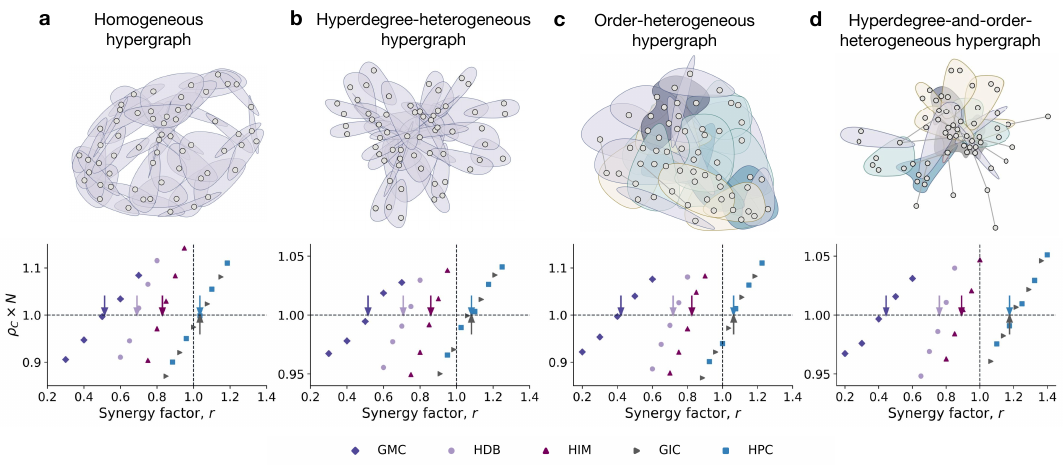}
		\caption{Numerical validations on homogeneous or heterogeneous hypergraphs. 
			The first row visualizes various hypergraphs:
			\textbf{a}, A homogeneous hypergraph with uniform hyperdegree and order; 
			\textbf{b}, A hyperdegree-heterogeneous hypergraph with the same order but varying hyperdegrees following a power-law distribution;
			\textbf{c}, An order-heterogeneous hypergraph with the same hyperdegree but varying orders following a Poisson distribution; 
			\textbf{d}, A hyperdegree-and-order-heterogeneous hypergraph with varying hyperdegrees and orders. 
			The second row compares theoretical results (arrows) with simulation data (scatters) on these hypergraphs. Each dot represents the fixation probability times population size from \(5 \times 10^5\) independent simulations under weak selection \(\delta= 0.025\). The arrows indicate the theoretical results for critical synergy factors $r^{\ast}$ listed in Table 1.
		}\label{fig:side}
	\end{figure*}
	
	\clearpage
	\begin{figure}[t]
		\centering
		\includegraphics[width=0.5\textwidth]{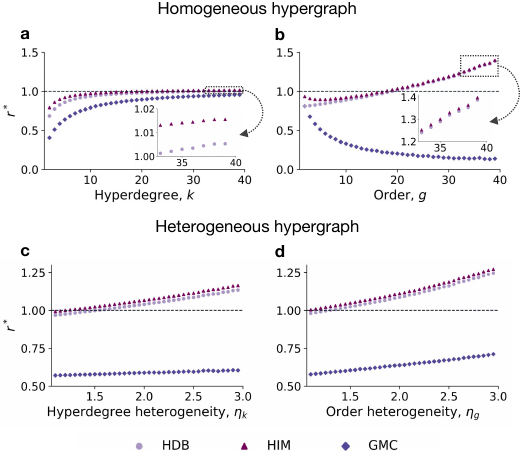}
		\caption{
			Impact of varying hyperdegrees and orders.
			\textbf{a}-\textbf{b}, Scatter plots of the critical synergy factor versus hyperdegree $k$ (\textbf{a}) and order $g$ (\textbf{b}) in homogeneous hypergraphs, where the hypergraphs have a size of 100, with a fixed order of 4 in \textbf{a} and a fixed hyperdegree of 4 in \textbf{b}.
			\textbf{c}-\textbf{d}, Scatter plots of the critical synergy factor versus hyperdegree heterogeneity $\eta_k$ in order-heterogeneous hypergraphs (\textbf{c}) and order heterogeneity $\eta_g$ in hyperdegree-heterogeneous hypergraphs (\textbf{d}), where the hypergraphs have a size of 100, with an average order of 8 and an average hyperdegree of 8 in \textbf{c} and \textbf{d}. Here, $\eta_k = \langle k^2 \rangle / \langle k \rangle^2$, where $\langle k^2 \rangle$ is the second moment of the hyperdegree, and $\eta_g = \langle g^2 \rangle / \langle g \rangle^2$, where $\langle g^2 \rangle$ is the second moment of the order.
			The scatter points in all plots are based on numerical calculations according to Table 1, on hypergraphs with corresponding topological configurations. All hypergraphs are constructed using the configuration model based on the given hyperdegree and order sequences.
		}
		\label{impact}
	\end{figure}
	\clearpage
	
	
	\begin{figure*}[t!]
		\centering
		\includegraphics[width=\textwidth]{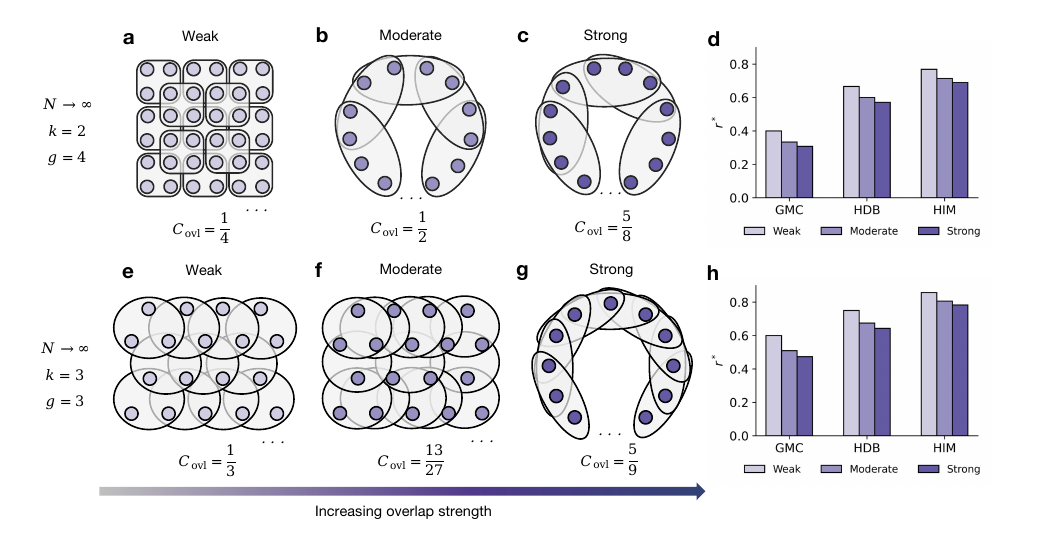}
		\caption{Increasing overlap strength can foster cooperation.
			\textbf{a}-\textbf{c}, Three homogeneous hypergraphs with large size $N$, hyperdegree $k=2$, and order $g=4$, differing only in the overlap strength between hyperedges, as defined in Eq.~\ref{overlap-homo}. The overlap strengths are $1/4$ in \textbf{a} (weak), $1/2$ in \textbf{b} (moderate), and $5/8$ in \textbf{c} (strong).
			\textbf{d}, Comparison of critical synergy factors for different overlap strengths under GMC, HDB, and HIM mechanisms.
			\textbf{e}-\textbf{g}, Another set of homogeneous hypergraphs with large size $N$, hyperdegree $k=3$, and order $g=3$, also differing in overlap strength: $1/3$ in \textbf{e} (weak), $13/27$ in \textbf{f} (moderate), and $5/9$ in \textbf{g} (strong).
			\textbf{h}, Comparison of critical synergy factors for increasing overlap strengths under the same mechanisms as in \textbf{d}.
			Both comparison results (\textbf{d}, \textbf{h}) show that stronger overlap decreases critical synergy factors across all the three mechanisms thereby promoting the emergence of cooperation.
		}\label{fig:side}
	\end{figure*}
	\clearpage

	\begin{figure*}[t!]
		\centering
		\includegraphics[width=\textwidth]{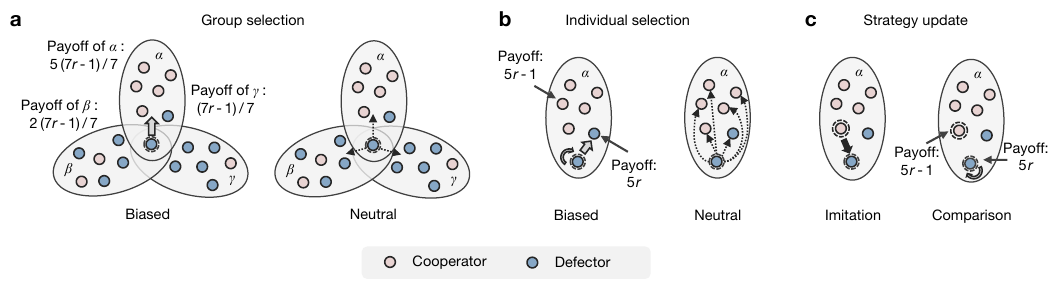}
		\caption{Intuitions from a viewpoint of defectors.
			Each individual (node) plays as a cooperator (pink) or a defector (blue) and participates in public goods games (PGGs) within the groups they belong to. After a game round, both the groups and the individuals receive payoffs, based on which the focal individual (a defector) updates its strategy by selecting an individual for imitation or comparison through two-stage selection. For simplicity, the group payoff is assumed to be the average of its individuals' payoffs and the individual payoff is the game outcome within a specific group.
			\textbf{a}, Group selection can be either group-biased or group-neutral. The focal player favors group $\alpha$ due to its higher payoff under group-biased selection, whereas it randomly chooses among neighboring hyperedges under group-neutral selection.
			\textbf{b}, Individual selection can be individual-biased or individual-neutral. The focal player prefers the defector, as defectors generally outperform cooperators in individual payoffs within a group under individual-biased selection, while it randomly picks a member within the group under individual-neutral selection.
			\textbf{c}, Strategy updates occur based on imitation or comparison. Imitation involves copying a preferential individual's strategy, whereas comparison tends to reinforce the original defector strategy due to the higher payoff itself.
		}\label{intuition}
	\end{figure*}
	\clearpage
	
	\begin{figure*}[t!]
		\centering
		\includegraphics[width=\textwidth]{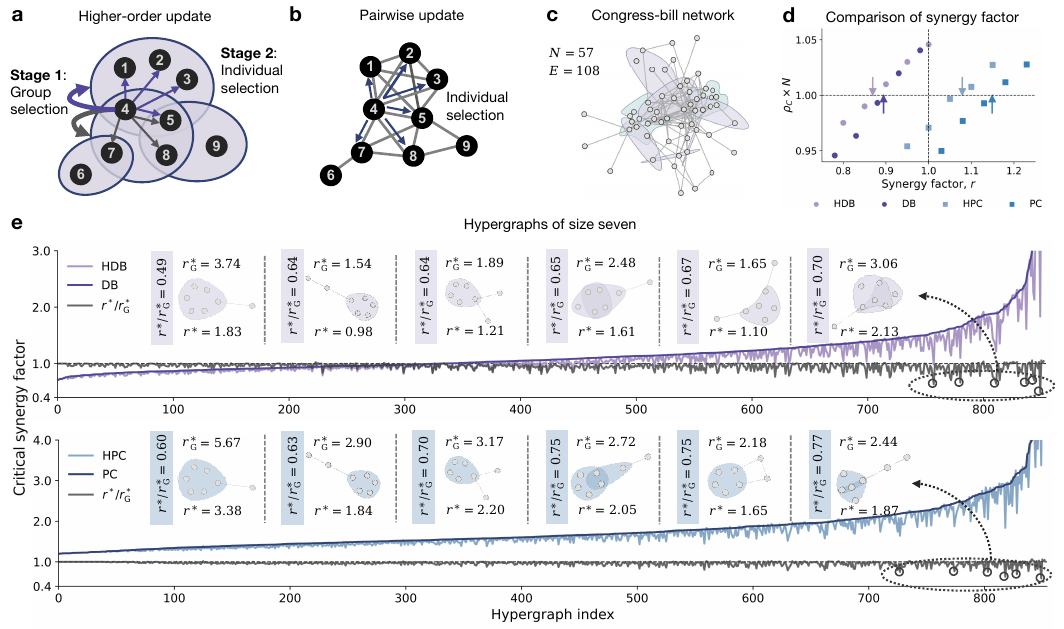}
		\caption{Higher-order updates tend to be more beneficial to cooperation than pairwise updates.
			\textbf{a}, Higher-order updates involve two-stage selection: group selection followed by individual selection.
			\textbf{b}, Pairwise updates involve only individual selection.
			\textbf{c}, An empirical higher-order network of congress bills with $N=57$ nodes and $E=108$ edges (or hyperedges), where each node represents a congressperson and each hyperedge represents a bill composed of sponsors and co-sponsors.
			\textbf{d}, Comparisons of synergy factors between higher-order updates (HDB and HPC) and pairwise updates (DB and PC).
			\textbf{e}, Comparisons between the critical synergy factor $r^{\ast}$ for higher-order updates and $r_{\text{G}}^{\ast}$ for pairwise updates. All seven-node subgraphs of the empirical network in \textbf{c} are extracted and sorted by $r_{\text{G}}^{\ast}$, with hyperedges formed by cliques. The results show that $r^{\ast} < r_{\text{G}}^{\ast}$ in most of the seven-node hypergraphs, indicating that higher-order updates generally promote cooperation more effectively than pairwise updates.
		}\label{fig:side}
	\end{figure*}
	\clearpage
	
	\begingroup
	\bibliographystyle{naturemag} 
	\bibliography{ref-sample}
	\endgroup
\end{document}